\pdfoutput=1
\documentclass[journal]{IEEEtran}
\usepackage{cite}
\usepackage{amsmath,amssymb,amsfonts}
\usepackage{graphicx}
\usepackage{booktabs}
\usepackage{siunitx}
\usepackage{textcomp}
\usepackage{xcolor}
\usepackage{multirow}
\usepackage{array}
\usepackage{tabularx}
\newcolumntype{Y}{>{\raggedright\arraybackslash}X}
\usepackage{url}

\usepackage{tikz}
\usetikzlibrary{positioning,fit,shapes.geometric}
\graphicspath{{../figures/}}

\begin{document}

\sloppy

\title{Measuring Control-Plane Openness in Near-Term Quantum Computing:\\
A Rubric, Its Validation, and an Application to Thirteen Vendor Stacks}

\author{\IEEEauthorblockN{Rylan N. Malarchick}
\IEEEauthorblockA{Department of Physical Sciences\\
Embry--Riddle Aeronautical University\\
Daytona Beach, FL, USA\\
malarchr@my.erau.edu}
\thanks{During the grading period the author was developing an open-source
pulse-control software project that used one of the vendor stacks surveyed here
(IQM) as its primary integration target; that project has since been
discontinued.  No claim in this paper depends on or references it, and the
IQM row is graded only on public documentation.}}

\maketitle

\begin{abstract}
Public access to pulse-level and control-electronics interfaces in commercial
quantum computing has bifurcated.  This paper proposes a six-axis rubric for
measuring control-plane openness, the layer between gate-level circuit
specification and physical control electronics, defined operationally so that
the same evidence produces the same grade across vendors.  The rubric is
validated three ways: a blinded re-grading pass that tests whether the cited
evidence and the level definitions alone reproduce the recorded grades, a
boundary-case methodology that fixes where each level begins and ends, and a
published grading protocol that lets others reproduce and contest any cell.
A time-point comparison anchored on the February 2025 removal of pulse-level
access from IBM hardware establishes that the rubric measures change rather
than describing a snapshot.  The rubric is applied to thirteen commercial
vendors across superconducting, trapped-ion, neutral-atom, and photonic
modalities as of May 1, 2026, and one of the three harms it detects is
demonstrated through a reproduction-access audit of five pre-2025 IBM Qiskit
Pulse experiments, carried through to a structural port to Rigetti Quil-T.
The catalog ships as a machine-readable artifact under CC-BY-4.0 with
per-cell source URLs.  The readings will go stale; the rubric is the
contribution that survives them.
\end{abstract}


\section{Introduction}
\label{sec:intro}

Public access to the control plane of commercial quantum computers has
bifurcated.  The largest superconducting cloud platforms have closed access at
this layer over the past three years, with IBM's removal of pulse-level
control from all production QPUs on February 3, 2025 standing as the most
consequential single event in the period under
review~\cite{ibmPulseMigration,qiskitDeprecate13063,qiskitRemove13662};
mid-tier superconducting vendors and the more open neutral-atom platforms have
moved in the opposite direction.  This paper proposes a rubric for measuring
that bifurcation, validates the rubric, and applies it to thirteen commercial
vendors as the rubric's first application.  The contribution is the
instrument.  The catalog is what the instrument reads on one date; the
instrument is what survives the readings going stale.

The control plane is the layer of the quantum computing stack between
gate-level circuit specification and the physical control electronics.  It
includes pulse synthesis, calibration loading and drift compensation, pulse
and measurement scheduling, measurement readout, and feedback dispatch.  It is
what sits between ``I want a CNOT'' and ``the FPGA emits these specific
microwave envelopes,'' and it is the layer at which most of what makes a
quantum experiment reproducible or hardware-aware actually lives.  The term is
borrowed from networking and cloud infrastructure and is not standard
quantum-computing terminology; we use it to refer specifically to this layer
for the duration of the paper.

The rubric grades the control plane on two surfaces.  The first is what
happens inside the control plane: pulse synthesis algorithms, scheduling
logic, feedback dispatch protocols, and the structure of the calibration
routine.  The second is what the control plane publishes outward to users:
calibration data, provenance metadata, reproducibility commitments, and the
schemas through which any of these are exposed.  A vendor can be open on the
inward surface and closed on the outward, or open on the outward and closed on
the inward, and the rubric distinguishes the surfaces in its axis structure.

\begin{figure}[t]
  \centering
  \begin{tikzpicture}[
    every node/.style={font=\footnotesize},
    layer/.style={
      draw,
      rounded corners=2pt,
      minimum width=0.92\columnwidth,
      minimum height=7.0mm,
      align=center,
      inner sep=2pt
    },
    cp/.style={layer, fill=black!8},
    nb/.style={layer, fill=white}
  ]
    \node[nb] (frontend) {Compiler frontend (gate-level circuits, OpenQASM 3, etc.)};
    \node[cp, below=1.5mm of frontend] (cp) {%
      \textbf{Control plane}\\
      pulse synthesis, calibration loading, scheduling,\\
      measurement readout, feedback dispatch};
    \node[nb, below=1.5mm of cp] (elec) {Control electronics (FPGA gateware, RF chain, AWGs)};
    \node[nb, below=1.5mm of elec] (hw)   {Physical hardware (qubits, lasers, traps, photonics)};
  \end{tikzpicture}
  \caption{The layers of a quantum computing platform as referenced in
  this paper.  The control plane (shaded) is the layer the rubric grades.
  The compiler frontend, control electronics, and physical hardware are out
  of scope for the rubric itself but are referenced in the per-axis
  definitions and in the limitations section to delineate what the control
  plane is and is not.  The figure is descriptive of what every commercial
  platform has, not prescriptive of any particular project's internal
  architecture.}
  \label{fig:stack}
\end{figure}
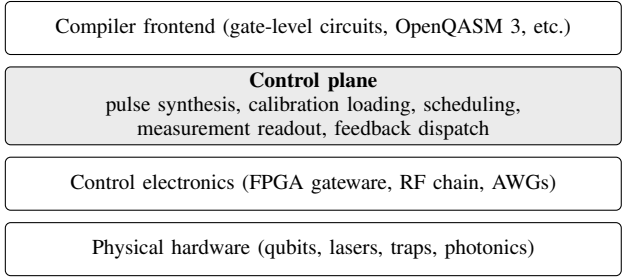

Fig.~\ref{fig:stack} shows the layers we have in mind.  This is the only
architectural figure in the paper.  We keep it as a flat horizontal stack
rather than an architectural decomposition: the layer labels describe what
every quantum computing platform has, not what any particular project's
internals look like.

\subsection{Why Measurement: Three Harms the Rubric Detects}
\label{sec:harms}

The bifurcation matters because three specific harms follow from it, and each
is a thing the rubric is built to detect.  We define each narrowly so that the
catalog and the demonstration in Section~\ref{sec:demo} can show it concretely
rather than assert it.

The first harm is \textbf{reproducibility}.  A previously published result
cannot be re-run on the platform it was performed on because the access it
required is no longer available.  The canonical example is the body of
pre-2025 IBM Quantum methods literature that used Qiskit Pulse: as of
February 3, 2025, IBM removed pulse-level control from all
QPUs~\cite{ibmPulseMigration}, which means a non-trivial fraction of that
literature cannot be re-executed on IBM hardware as of the May 1, 2026 cutoff
used here.  This is not a hypothetical harm; it is dated, and the rubric's
column on pulse-level access is what pins which results are affected.

The second harm is \textbf{hardware-aware research}.  A class of experiments
that depends on access to a specific layer cannot be performed on a given
platform because the access is not exposed.  Examples include closed-loop
measurement-conditioned feedback, which depends on real-time pulse dispatch at
the control electronics; novel pulse decompositions for specific gate sets,
which depend on raw waveform synthesis; dynamical decoupling sequences, which
depend on precise inter-pulse spacing; and calibration drift studies, which
depend on published calibration data over
time~\cite{wisemanMilburn2009,closedLoopFeedback2022}.  The rubric records,
per platform, which of these are publicly possible.

The third harm is \textbf{cross-vendor benchmarking}.  The same benchmark
protocol cannot be executed identically on two platforms because the
calibration metadata is published in incompatible formats or not at all, or
because the underlying pulse-level interface differs in ways that make ``the
same circuit'' mean different things across hardware.  Comparing a closed-loop
benchmark on Rigetti, which exposes Quil-T pulse-level access, to one on IonQ,
which exposes native gates only, is structurally impossible at the pulse
layer; the comparison can only happen at a layer of abstraction high enough to
lose the signal the benchmark is trying to
measure~\cite{rigettiQuilt,ionqNative,qedcBenchmarks,benchpress}.

This paper does not propose an architecture or a reference implementation.  It
proposes an instrument for measuring control-plane openness, validates it,
applies it to the commercial landscape, and identifies what minimally open
access at this layer would have to look like.

\subsection{Paper Structure}

Section~\ref{sec:rubric} defines the rubric axis by axis.
Section~\ref{sec:validation} validates it through a blinded re-grading
stability test, a boundary-case methodology, and a time-point sensitivity
analysis.
Section~\ref{sec:scope} names what is in and out of scope and the grading
procedure.  Section~\ref{sec:related} positions the rubric relative to
openness measurement in adjacent ecosystems, quantum software perspectives,
and the quantum reproducibility literature.  Section~\ref{sec:catalog} applies
the rubric to thirteen vendors.  Section~\ref{sec:results} reads what the
application shows.  Section~\ref{sec:demo} demonstrates one of the three harms
through a concrete reproduction-access audit.  Section~\ref{sec:steelman}
treats eight alternative interpretations of the readings.
Section~\ref{sec:analogues} draws lessons from three open-infrastructure
precedents.  Section~\ref{sec:floor} identifies what minimally open access
would require.  Section~\ref{sec:conclusion} concludes.  The catalog itself is
released as a separate machine-readable artifact with source URLs and
accessed-on dates per cell.

\section{The Control-Plane Openness Rubric}
\label{sec:rubric}

The rubric is the paper's primary contribution.  It defines six axes of
control-plane openness, each with discrete levels and explicit grading
criteria, so that the same public evidence produces the same grade
independent of who applies it.  Each axis below states (a) what it measures,
(b) why it is part of openness as opposed to general software quality, (c) the
levels with their grading criteria, and (d) the boundary case that drove where
one level ends and the next begins.  The boundary cases are not incidental;
they are the part of the rubric that makes the levels reproducible, and
Section~\ref{sec:validation} treats their precision as one leg of validation.

\subsection{Axis 1: SDK Pulse-Level Access}
\label{sec:axis1}

This axis measures whether public-API users can specify the shape of the
control signal, not merely its label.  It is part of openness rather than
software quality because waveform-level access is the interface at which a
published pulse-level result becomes independently executable; an SDK that
abstracts the waveform away can be excellent software and still closed at the
layer the result lived in.  \emph{Open} requires raw waveform synthesis exposed
to public-API users, including custom envelopes, frame manipulation, and
sample-level control.  \emph{Partial} requires parameterized waveform templates
only, without sample-level edit.  \emph{Native-gate-only} requires the native
gate set exposed via API but no waveform manipulation.  \emph{Closed} requires
no public pulse or native-gate API beyond standard gate-level circuits.  The
boundary that drove the level definitions sits between partial and
native-gate-only: a vendor exposing parameterized DRAG templates with editable
amplitude and duration but no sample array grades partial, while a vendor
exposing a fixed native gate with a phase argument grades native-gate-only.
The discriminating question is whether the user can change the shape of the
emitted envelope, not merely a parameter of a fixed shape.  Two further
boundaries on this axis were fixed by the re-grading pass of
Section~\ref{sec:regrading}.  A native gate set that is publicly documented
but whose hardware execution is not publicly accessible grades closed rather
than native-gate-only; the axis grades public access, not public
documentation of access someone else has.  And parameterized basis-gate
instructions inside a standard circuit interface do not constitute a
native-gate API; native-gate-only requires a direct native-gate submission
path with documented pulse-level phase semantics, not a richer transpilation
basis.

\subsection{Axis 2: Calibration Data Publication}
\label{sec:axis2}

This axis measures whether the device state that produced a result is
publicly obtainable.  It is part of openness because a result's numerical
values are uninterpretable without that state; withholding it does not lower
software quality, it removes the context a reader needs to know what the
numbers mean.  \emph{Open} requires per-device calibration data published on a
documented cadence in a documented schema, with retroactive access.
\emph{Partial} requires per-job calibration metrics fetchable but no historical
archive, or summary statistics only.  \emph{Limited} requires calibration data
available on request or to research partners under separate agreements.
\emph{Closed} requires no calibration data published or fetchable.  The boundary
that drove the definitions sits between partial and limited: data an arbitrary
public user can fetch through a documented API call grades partial; data
released only to named partners under a bilateral agreement grades limited.

\subsection{Axis 3: Control Electronics Interface}
\label{sec:axis3}

This axis measures whether the binding between SDK and physical electronics is
a documented contract.  It is part of openness because reproducing a
closed-loop or feedback result requires knowing what physical loop closed and
when; an undocumented coupling between SDK and gateware is an opaque dependency
even when every layer above it is open source.  \emph{Open} requires the
SDK-to-hardware schema documented, third-party control electronics
substitutable, and FPGA gateware between measurement and control documented or
open-source.  \emph{Partial} requires one of those three documented but not all.
\emph{Limited} applies when the vendor uses standard commercial control
electronics (Quantum Machines OPX, Zurich Instruments) without publishing the
integration layer.  \emph{Closed} requires no public information on the
control-electronics layer.  The boundary that drove the definitions sits
between partial and limited: documenting any one sub-criterion grades partial,
whereas using knowable commercial electronics without publishing the binding
grades limited, because the components are identifiable but the contract is
not.

\subsection{Axis 4: Queue Model}
\label{sec:axis4}

This axis measures whether the scheduling model permits real-time and feedback
work.  It is part of openness because closed-loop experiments are impossible
without scheduling access that a shared best-effort queue does not provide; the
constraint is structural, not a matter of convenience.  \emph{Open} requires
dedicated reservation or on-prem deployment available, with real-time loops
possible and scheduling guarantees documented.  \emph{Partial} applies to mixed
access, where general users get a shared queue and a reservation tier exists
for real-time work.  \emph{Limited} applies to shared cloud queue only, with no
real-time guarantees and latency that varies by load.  \emph{Closed} applies
when no public information on queue or scheduling model is available.  The
boundary that drove the definitions sits between partial and limited: a
documented reservation tier alongside the shared queue grades partial; a shared
cloud queue with no real-time path grades limited.

\subsection{Axis 5: Reproducibility Guarantees}
\label{sec:axis5}

This axis measures whether returned results can be situated in time.  It is
part of openness because a result without a calibration snapshot identifier, a
hardware revision, and an SDK version cannot be placed against any other run;
the metadata is what lets a later reader establish that two results are or are
not comparable.  \emph{Open} requires calibration versioned with stable
identifiers, returned results carrying provenance metadata, and hardware
revision stability commitments published.  \emph{Partial} requires versioned
calibration but no provenance metadata, or provenance present but no stability
commitments.  \emph{Limited} requires a best-effort posture documented.
\emph{Closed} applies when no reproducibility-related metadata is returned with
jobs and no commitments are published.  The boundary that drove the definitions
sits between open and partial: all three of versioned calibration, provenance
metadata, and a published stability commitment grades open; any two of the
three grades partial.

\subsection{Axis 6: SDK License}
\label{sec:axis6}

This axis measures whether the software layers above the hardware can be
inspected.  License is the part of openness most often mistaken for the whole
of it.  It is graded separately precisely so that the catalog can show a
permissively licensed SDK sitting on top of a closed control plane, the IBM
case, without letting the license stand in for access.  \emph{Open} requires
Apache 2.0, MIT, BSD-3-Clause, or another OSI-approved permissive license,
with full source available.  \emph{Partial} requires a permissive license on the
public layer but proprietary or closed components below it.  \emph{Limited}
requires source available under a non-commercial or restricted-use license.
\emph{Closed} applies when the SDK is proprietary, has no public source, or does
not exist.  The boundary that drove the definitions sits between partial and
limited: a permissive license on the public layer with proprietary components
below grades partial; source under a non-commercial or restricted-use license
grades limited.

\subsection{Axis Independence}
\label{sec:independence}

The six axes are graded independently: a vendor can score open on one and
closed on another, and the catalog records several such combinations.  IBM is
open on license and closed on pulse access; Atom Computing is closed on four
axes and limited on the remaining two.  Independence is an assumption the rubric makes
and a property the catalog can check.  Section~\ref{sec:sensitivity} reports
whether the axes move together or separately across the time-point comparison,
which is the empirical test of whether they measure distinct things rather
than restating a single underlying score.

\section{Validation}
\label{sec:validation}

A rubric that is never validated is a list of opinions with headings.  This
section establishes that the rubric measures what it claims, applies
consistently, and detects change.  Validation proceeds in three parts:
construct validity (what the rubric measures and why that is openness),
re-grading stability (whether the same evidence produces the same grade on
re-application), and a time-point sensitivity analysis (whether the rubric
detects change rather than describing a static snapshot).  The boundary-case
definitions in Section~\ref{sec:rubric} are the operational-precision leg that
makes the first two possible.

\subsection{Construct Validity}
\label{sec:construct}

Construct validity asks whether the rubric measures the thing its name claims.
The construct is control-plane openness, defined operationally as the degree to
which the interfaces between gate-level specification and physical control are
publicly accessible, publicly documented, and exposed with the metadata a
reader needs to reproduce a result.  This is narrower than openness in general,
and the narrowing is deliberate.  The rubric does not measure license openness
alone (Axis 6 is one axis of six, and the IBM row exists to show why license
cannot stand in for the rest), community openness (whether a project accepts
outside contributions), or governance openness (who controls the roadmap).
The Open Source Definition fixes the license-and-redistribution sense of the
word~\cite{osiOSD}; the construct here borrows the public-accessibility
intuition but applies it to runtime access rather than to source
redistribution.  Bova and Melko argue the field-level case that openness in
quantum is more complementary to commercial interest than competitive with
it~\cite{bovaMelko2025}; the rubric operationalizes the specific slice of
openness that argument depends on, the part a researcher needs in order to
reproduce and extend a hardware result.  Naming what the rubric excludes is
itself part of the validity claim: an instrument that measured all of openness
at once would measure none of it precisely.

\subsection{Re-Grading Stability}
\label{sec:regrading}

Re-grading stability tests whether the rubric is precise enough that a grader
applying it to the same evidence, with no knowledge of the recorded grades,
produces the same grades.  An instrument that returns different readings on
re-application is recording the grader's state rather than the vendor's.  The
catalog was first graded in late April 2026 against sources accessed on or
before the May 1 cutoff.  On June 9, 2026, thirty-nine days after the cutoff,
the catalog was re-graded under a masked protocol: a script stripped every
recorded grade, cell text, rationale, and commentary from the catalog notes,
leaving the rubric's level definitions, each cell's cited URL with its
accessed-on date, and the verbatim source quotes recorded at first grading.
The re-grade was performed against this pack and live fetches of the cited
URLs by an AI assistant in a fresh session with no access to the recorded
grades or any other project file (see the Tool Use Disclosure), under
instructions to return a level only where the evidence decides one, to return
``insufficient evidence'' otherwise, and to consult nothing beyond the listed
sources.  The masking is structural rather than memory-based: the re-grader
could not recall the original levels because it never saw them.  The pack,
the re-grade output, and the comparison script ship with the catalog
artifact, so the pass can be audited or repeated by anyone, with any grader.

Of the 78 cells, 8 are recorded as ``see notes'' and are excluded, leaving
70.  The re-grade reproduced the recorded level exactly on 29 cells, returned
a different level on 6, and returned insufficient evidence on 35.  The three
numbers say different things about the instrument, and reading them requires
keeping them apart.

The 35 insufficient-evidence cells are failures of the evidence locus, not of
the level definitions: on these cells the re-grader did not weigh the same
evidence and reach a different conclusion; it found that the recorded
citation no longer carries, or never alone carried, the cell.  Sixteen of the
35 cited pages were dead, redirected off their recorded host, or returned no
machine-readable content at re-grade time, thirty-nine days after the
evidence cutoff: the Quantinuum hardware guide had moved (a break already
noted at first grading), the IQM documentation host now redirects to a
successor corporate domain (iqm.tech) adopted during the vendor's commercial
transition in the same window, strawberryfields.ai redirects to a corporate
page that no longer describes the SDK, and the cited AWS Braket device page
no longer lists the vendor.  The content often persists somewhere; the recorded
citation no longer resolves to it.  The remaining 19 pages were live but do
not by themselves state the cell's grading basis, which records a
citation-granularity defect in the artifact: first grading drew on
vendor-wide source sets, and the artifact records one URL per cell.  Both
failure modes concentrate on Axes 2, 4, and 5 (nine, nine, and ten of the
35), the axes whose evidence lives in operational documentation, consoles,
and account-holder material rather than in stable repository or specification
pages.  Axis 5 was one of the two predicted disagreement sites, and the
prediction is confirmed in an unexpected form: all ten of its disagreements
are evidence failures, none is a level disagreement.  Axis 3, the other
predicted site, largely held (four evidence failures, no level
disagreements), because on the electronics axis the documented absence of
public information is itself gradable evidence.  There is an irony here the
paper accepts rather than hides: the catalog's own evidence layer decays at a
measurable rate, which is the same reproducibility failure at the
documentation surface that the rubric measures at the access surface.  The
insufficiency log ships with the artifact, and per-cell citation granularity
is named in the artifact's changelog as the first correction target for the
next catalog iteration.

Among the 35 cells where the cited evidence still decided a level, 29
reproduced exactly (83 percent).  The six level disagreements are the
load-bearing validation data, and each was adjudicated against the rubric
text after unmasking.  Two sit on the same underspecified Axis 1 boundary and
forced the refinements now recorded in Section~\ref{sec:axis1}: the Google
cell (recorded closed, re-graded native-gate-only) turns on a documented gate
set whose hardware execution is available only through sponsored
collaboration, and the post-2025 IBM cell (recorded closed, re-graded
native-gate-only) turns on whether fractional-gate instructions inside the
standard circuit interface constitute a native-gate API.  Under the refined
boundary, both recorded grades stand.  One disagreement caught a boundary
misapplication, and the catalog is corrected for it: the Google calibration
cell (recorded partial, re-graded limited) fails the Axis 2 boundary's own
discriminating question, because an arbitrary public user cannot invoke the
documented calibration query without a sponsored Engine project; the cell now
grades limited, and the correction is dated in the artifact.  The
disagreement pattern on Axis 4 prompted a rubric-consistency sweep of that
axis, which found one further misgrade: the ORCA queue cell recorded partial
where no level fits institutional-partnership access with no public queue,
and it is corrected to ``see notes'' per the grading procedure of
Section~\ref{sec:scope}.  The remaining three disagreements stand as logged
disagreements rather than corrections, and two of them are citation defects
rather than grading defects.  The Quantinuum license cell (recorded partial,
re-graded open) rests on a proprietary direct client alongside the Apache-2.0
compiler, documented in the catalog notes but not visible from the cell's
single cited URL.  The QuEra queue cell (recorded partial, re-graded limited)
rests on the cloud provider's documented reservation tier, available for all
hosted devices alongside the shared queue, which the recorded citation does
not mention; the supporting source is now recorded in the notes with a dated
repair.  The Xanadu pulse cell (recorded open, re-graded closed) is the one
substantive disagreement the surviving public evidence cannot adjudicate: the cited host now redirects, and the question underneath, whether
continuous-variable gate parameters such as squeezing magnitudes and
beamsplitter angles constitute analog control of the emitted signal, is a
modality-translation question on which the rubric's superconducting-shaped
vocabulary is genuinely ambiguous.  It is logged in the artifact as
contested, and it marks the rubric's clearest current limitation for photonic
rows.

After the corrections, the catalog stands at 69 gradable cells and 9 ``see
notes.''  Reporting the decomposition rather than a single agreement number
is the point.  The level definitions held where the evidence held; the
dominant instability is the evidence layer, not the grader; and the two cells
that moved, moved because the re-grade caught the rubric being misapplied,
which is the failure mode this validation exists to catch.  The rubric is
offered as an instrument others can apply, and the decomposition is the
honest estimate of how repeatable that application is.

\subsection{Time-Point Sensitivity}
\label{sec:sensitivity}

A measurement instrument must detect change, not merely describe a state.  If
the rubric returns the same grades regardless of when a vendor's documentation
is read, it is describing a category rather than measuring a quantity that
moves.  The window for the comparison is anchored before the largest
control-plane event of the period, the removal of pulse-level access from IBM
Quantum hardware on February 3, 2025, and runs from late 2024 to the May 1,
2026 primary cutoff.  A window ending in mid-2025 would open after the IBM
transition and miss it, so the twelve-months-ending-at-cutoff framing is not
the right one here.  Each candidate change is established from dated primary
sources, release notes, deprecation issues, and vendor announcements, rather
than from re-grading archived documentation, because archived page bodies were
not reliably retrievable at grading time; cells for which no dated change was
found are reported as showing no documented movement, not as confirmed stable.

The headline delta is the IBM Axis 1 cell.  Through 2024, IBM exposed
sample-level pulse control to public-API users through Qiskit Pulse and
OpenPulse, which grades open.  Backend support for Qiskit Pulse instructions
was removed from all IBM QPUs on February 3, 2025~\cite{ibmPulseMigration,
qiskit13release}, and the \texttt{qiskit.pulse} module was removed from the SDK
in Qiskit 2.0~\cite{qiskitRemove13662,qiskitDeprecate13063}; as of the cutoff
the cell grades closed.  The documented migration target is fractional gates
plus Qiskit Dynamics, neither of which restores sample-level waveform synthesis
to public-API users.  This single cell moving from open to closed, dated to a
specific day and forced by IBM's own documentation rather than by the grader's
judgment, is the clearest evidence that the rubric registers
measurement-relevant change.  A rubric on which that cell did not move would be
failing to register the most consequential control-plane event of the period.

\begin{table}[t]
\caption{Time-point deltas established from dated primary sources, anchored
before the February 3, 2025 removal of IBM pulse access and read at the
May 1, 2026 cutoff.  The IBM Axis 6 row is the axis-independence control: it
held open while Axis 1 closed.}
\label{tab:delta}
\centering
\small
\setlength{\tabcolsep}{3.5pt}
\begin{tabular}{@{}llccl@{}}
\toprule
Vendor & Axis & Before & Cutoff & Evidence \\
\midrule
IBM & 1 (pulse) & open & closed & Feb 3 2025 removal~\cite{ibmPulseMigration} \\
IBM & 6 (license) & open & open & Qiskit Apache 2.0 throughout \\
IQM & 1 (pulse) & \multicolumn{2}{c}{candidate} & Pulla maturation~\cite{iqmPullaBlog} \\
OQC & 1 (pulse) & open & open & open since Oct 2022~\cite{oqcOpenPulse} \\
\bottomrule
\end{tabular}
\end{table}

The same window tests axis independence.  IBM Axis 6 (SDK license) did not move:
Qiskit has been Apache 2.0 throughout and remains open at the cutoff.  A vendor
whose pulse-access axis closes while its license axis holds is the concrete case
the independence assumption predicts; if the six axes were restating one
underlying score, Axis 6 would have tracked Axis 1 downward, and it did not.

Two further changes were examined and are reported as candidates rather than
confirmed deltas.  IQM's public pulse-level offering, Pulla, matured into a
documented product across the window~\cite{iqmPullaBlog}, consistent with the
IQM Axis 1 cell strengthening toward open; a clean transition date for public
access, as opposed to partner or on-premises access, could not be fixed from the
primary sources, so the cell is not entered as a dated delta.  OQC pulse-level
access, by contrast, has been public since October 2022~\cite{oqcOpenPulse} and
did not move within the window.  Across the remaining vendors no dated grade
change was found in the sources reviewed; because archived page bodies could not
be re-graded cell by cell, this is recorded as no documented movement rather
than confirmed stability, and a full archived re-grade of every cell remains
available as a way to strengthen the analysis.

\section{Scope and Methodology}
\label{sec:scope}

This paper covers gate-model commercial quantum computing platforms across
four hardware modalities: superconducting transmon, trapped ion, neutral atom,
and photonic.  The rubric is applied to thirteen vendors at the control plane
as defined in Section~\ref{sec:intro}, using public documentation accessed on
or before May 1, 2026.

The grading procedure is fixed and the same for every cell.  For each
vendor-axis pair: (a) the public documentation was accessed at a specific URL;
(b) the rubric-relevant claim was located within it; (c) the level definition
from Section~\ref{sec:rubric} was applied to that claim; and (d) the URL and
the accessed-on date were recorded.  The recorded URL and date are what make a
cell auditable and what let the time-point sensitivity analysis re-grade
against an archived version of the same source.  Cells that resist
categorization under the rubric use a ``see notes'' level with a brief
explanation in the per-row commentary.

We do not cover quantum error correction codes considered as algorithmic
objects, quantum networking, post-quantum cryptography, classical simulators of
quantum systems, or hardware modalities not yet in commercial deployment
(topological qubits, NV-center systems, color-center spin qubits, and others).
Open Quantum Design is treated as related work in Section~\ref{sec:related}
rather than as a catalog row, on the basis that as of the cutoff date OQD is a
foundation building reference designs rather than a commercial vendor running
jobs for outside users; the row could appear in a future iteration of the
catalog if that changes~\cite{oqdAnnouncement}.  QCi (Quantum Computing Inc) is
not included as a separate row because the company's Dirac systems are aimed
primarily at optimization workloads rather than gate-model quantum computing,
which puts them outside the rubric's axes.

The catalog grades public documentation only.  Where a vendor offers private
access to research partners under separate agreements, the catalog notes the
existence of the private access but does not grade on it; the paper is about
what is publicly citable, reproducible, and auditable, and the grading reflects
that.

\section{Related Work}
\label{sec:related}

Three threads of prior work inform this paper: openness measurement in
adjacent infrastructure ecosystems, perspectives on quantum software openness,
and the quantum reproducibility and benchmarking literature.

\subsection{Openness Measurement in Adjacent Ecosystems}

Measuring openness is an established practice in infrastructure layers older
than quantum computing.  The Open Source Definition gives the canonical
operational test for whether a license is open, reducing a contested word to a
checklist of redistribution and modification rights~\cite{osiOSD}; the rubric
here adopts the same move of pinning a contested word to discrete, checkable
criteria, but applies it to runtime access rather than to source code.  The
RISC-V instruction set architecture made the case that an open specification at
the bottom of a hardware stack changes who can build on it, and that the
openness in question is about the published interface rather than about any
particular implementation~\cite{riscvISA}.  The semiconductor design stack
offers the closest structural precedent: the SkyWater 130nm process design kit
opened a layer that had been NDA-locked, and subsequent open-EDA work
(OpenROAD and the tooling around it) measured and built on that openness at the
layer where the foundry's IP sat closest to the
designer~\cite{skywaterPDK,edwardsSkywater2020,openroad}.  These efforts share
a structure with the present work: openness is treated as a property of a
specific interface layer, measured against the public artifacts at that layer,
rather than as a vague disposition of a vendor.

\subsection{Quantum Software Ecosystem Perspectives}

The case for openness in quantum has been made at the level of the field; this
paper provides measurement.  Bova and Melko argue that an open-source
initiative would benefit quantum computing by reducing R\&D costs, improving
benchmarks, and growing the talent pool, and that openness in quantum is more
complementary to industry than competitive with it~\cite{bovaMelko2025}.  Open
Quantum Design is the institutional version of that argument running: a
non-profit founded in 2024 that announced an open-source, full-stack
trapped-ion quantum computer in January 2025, with partnerships across academia
and the open-source quantum community~\cite{oqdAnnouncement}.  The openQSE
reference architecture, drawn from a survey of nine production quantum-HPC
software stacks, proposes a vendor-neutral set of layer boundaries for runtime,
resource management, orchestration, and execution~\cite{openQSE2026}, and the
Munich Quantum Software Stack and its Quantum Device Management Interface
describe a hardware-software interface released under Apache 2.0 with LLVM
Exceptions, positioned as a de facto standard for hardware
abstraction~\cite{munichQDMI,munichMQSS}, with pulse-level support for that
stack the subject of active engineering work~\cite{mqssPulse2025}.  These
works argue for openness or
propose architectures that assume it; none measures the per-vendor access
landscape that an architecture would have to wrap or that the field-level
argument depends on.  The rubric supplies that measurement, and the catalog is
the dataset any such proposal has to engage with.

\subsection{Quantum Reproducibility and Benchmarking}

The harms the rubric is built to detect have been independently surfaced by the
reproducibility and benchmarking literature.  Benchmarking suites including the
QED-C application-oriented benchmarks and Benchpress depend on a stable,
documented relationship between a circuit specification and what the hardware
executes, which is exactly the relationship the control-plane axes
grade~\cite{qedcBenchmarks,benchpress}.  Work on reproducibility in quantum
software experiments has named the same access dependencies from the
reproducibility side: that re-executing a published quantum result requires the
calibration state, the hardware revision, and the interface access the original
used~\cite{repro123,reproBuilds,noisyRepro}.  This literature establishes that
the harms in Section~\ref{sec:harms} are recognized problems; the rubric's
contribution relative to it is to measure, per vendor, which access dependencies
are currently satisfiable in public, which is the precondition for knowing
whether a given result is reproducible at all.

\section{Application: Thirteen Vendor Stacks}
\label{sec:catalog}

The first application of the rubric grades thirteen vendors across four
hardware modalities on the six axes of Section~\ref{sec:rubric}.  Each cell
carries a level grade and a citation that justifies the level; cells that
resist categorization use a ``see notes'' level explained in the per-row
commentary.  The full machine-readable version of the table, including source
URLs and accessed-on dates per cell, ships as the Zenodo artifact under
CC-BY-4.0.  The version below is the printed companion.

\begin{table*}[t]
\caption{Control-Plane Openness Catalog (rubric applied as of May 1, 2026;
two cells, Google calibration and ORCA queue, corrected June 9, 2026
following the re-grading pass of Section~\ref{sec:regrading}, with the
corrections dated in the artifact)}
\label{tab:catalog}
\centering
\scriptsize
\setlength{\tabcolsep}{4pt}
\begin{tabularx}{\textwidth}{@{}clXccccc@{}}
\toprule
\# & Vendor & Modality & C1 SDK pulse & C2 Cal. data & C3 Electronics &
C4 Queue & C5 Repro. \\
\midrule
 1  & IBM Quantum       & Superconducting       & closed             & partial & closed  & partial & partial \\
 2  & Google Quantum AI & Superconducting       & closed             & limited & closed  & closed  & partial \\
 3  & IonQ              & Trapped ion           & native-gate-only   & partial & closed  & partial & partial \\
 4  & Quantinuum        & Trapped ion           & native-gate-only   & partial & closed  & partial & partial \\
 5  & IQM               & Superconducting       & open               & partial & partial & partial & partial \\
 6  & Rigetti           & Superconducting       & open               & partial & partial & partial & partial \\
 7  & Pasqal            & Neutral atom          & open               & partial & partial & partial & partial \\
 8  & QuEra             & Neutral atom          & open               & partial & see notes & partial & partial \\
 9  & Atom Computing    & Neutral atom          & closed             & closed  & closed  & closed  & limited \\
10  & OQC               & Superconducting       & open               & partial & see notes & partial & partial \\
11  & Xanadu            & Photonic (CV/GBS)     & open               & partial & see notes & partial & partial \\
12  & PsiQuantum        & Photonic              & closed             & closed  & closed  & closed  & closed \\
13  & ORCA Computing    & Photonic              & see notes          & see notes & see notes & see notes & see notes \\
\bottomrule
\end{tabularx}

\smallskip
SDK license (Axis 6): IBM open (Apache 2.0); Google open (Apache 2.0);
IonQ open (Apache 2.0 via Qiskit/Cirq/PennyLane integrations);
Quantinuum partial (TKET Apache 2.0, proprietary direct client);
IQM open; Rigetti open (Apache 2.0); Pasqal open (Apache 2.0);
QuEra open (Apache 2.0); Atom Computing limited (no public pulse SDK);
OQC open (via AWS Braket SDK, Apache 2.0); Xanadu open (Apache 2.0);
PsiQuantum closed; ORCA see notes (proprietary, institutional access only).
\end{table*}

The ``see notes'' level on row~13 (ORCA) reflects that the public surface for
grading SDK pulse-level access is smaller than the rubric anticipates: ORCA
exposes a PyTorch-based SDK with NVIDIA CUDA-Q integration, but the level of
pulse-level or analog-parameter control exposed through that SDK is not fully
documented in publicly accessible sources as of the cutoff~\cite{orcaTech}.
The queue cell joined the ``see notes'' set on June 9, 2026, when the
re-grading pass of Section~\ref{sec:regrading} prompted an Axis 4 consistency
sweep: no rubric level fits institutional-partnership access with no public
queue, and the grading procedure of Section~\ref{sec:scope} assigns ``see
notes'' to cells no level fits.

\subsection{Per-Row Commentary}
\label{sec:rowCommentary}

The per-row commentary follows a fixed three-sentence structure across all
thirteen rows.  Sentence one names the headline access status and the most
consequential single fact about the vendor's control plane.  Sentence two names
the axis-by-axis nuances the table cells alone do not convey.  Sentence three
names what makes the row interesting in the bifurcation story.  Rows whose
structural role is already covered by another row get a two-sentence commentary.

\textbf{Row 1: IBM Quantum.}  IBM is the load-bearing case in the catalog:
pulse-level control via Qiskit Pulse was deprecated in November 2024 and
removed from all production QPUs on February 3, 2025, with the replacement
(fractional gates plus Qiskit Dynamics simulation) covering the most common use
cases at a higher level of abstraction but not exposing raw waveforms, custom
envelopes, or frame manipulation to public-tier
users~\cite{ibmPulseMigration,qiskitDeprecate13063,qiskitRemove13662,qiskit13release}.
Calibration grades partial because per-job metrics remain fetchable through
\texttt{backend.properties()} but historical archives are not retained on a
documented cadence; the queue model reflects tiered access (open shared queue,
pay-as-you-go, and Network reservation) with real-time loops not generally
available on the open tier post-deprecation; the SDK license is Apache 2.0
throughout, but licensing alone does not prevent the access regression at the
runtime layer.  This row is the load-bearing case for the bifurcation argument:
IBM has the largest superconducting cloud user base, ran an open pulse-level
interface for several years, and chose to close it; whatever else the catalog
shows, the IBM row is the single most consequential data point about where the
trajectory has gone.

\textbf{Row 5: IQM.}  IQM Pulla is a client-side Python library distributed
through PyPI that exposes pulse-level access with full step-by-step compilation
transparency from circuit to pulse schedule, placing IQM among the most open
major-vendor stacks at the pulse layer~\cite{iqmPulla,iqmPullaBlog}.  Particular
care was taken on this row because of the disclosure footnote on page~1: every
cell is graded only on public IQM documentation, and the author's prior software
experience with the stack is excluded from the grading basis.  This row is the
counterexample to the simple ``big players close'' narrative: IQM is a
commercially funded superconducting vendor of non-trivial scale whose
pulse-level access trended open over the same period in which IBM's trended
closed.

\textbf{Row 7: Pasqal.}  Pulser is the cleanest open-by-design pulse-level
interface in the catalog: an Apache-2.0 Python package that exposes pulse-level
and analog control of neutral-atom devices with maximal user control over
physical parameters within device
bounds~\cite{pasqalPulser,pasqalPulserDocs}.  The calibration and
reproducibility cells are confirmed against Pasqal Cloud Services documentation.
This row, with QuEra, establishes that the neutral-atom modality is
open-by-design at the SDK and pulse layers, and it is the citable example for
the floor-section discussion of what ``open pulse APIs'' looks like in practice.

\textbf{Row 12: PsiQuantum.}  PsiQuantum is the cleanest closed-by-design case
in the catalog: the Omega chipset announced February 2025 is purpose-built for
fault-tolerant, million-qubit-scale operation, and the Construct platform exists
for fault-tolerant algorithm development but is not publicly distributed as an
SDK~\cite{psiquantumOmega,psiquantumConstruct}.  All six axes grade closed,
reflecting that the company has chosen not to expose intermediate-scale access
publicly while it builds toward fault tolerance.  This row is the structural
opposite of Xanadu in the photonic modality, and the contrast demonstrates that
modality alone does not predict openness: the same broad hardware substrate can
underpin maximally closed and maximally open commercial postures depending on
strategic position.

\textbf{Row 2: Google Quantum AI.}  Google Quantum AI has never exposed raw
pulse access on Sycamore through the public Cirq API; calibration metrics
including XEB error rates per gate are fetchable through
\texttt{cirq\_google.Engine} by collaboration users with a sponsored Engine
project (the calibration cell grades limited for that reason), and waveform
control is not exposed at all~\cite{googleCirqCalib,googleCirqDevices}.  Access is by research
collaboration rather than a public cloud queue, which is structurally different
from IBM's deprecation path: Google was never open at the public pulse layer to
begin with, and the grades reflect that initial-state closure rather than a
trajectory away from openness.  This row is the second closed-by-default
superconducting case and shows that ``no public access'' is a position the field
encounters from the start, not only after a deprecation.

\textbf{Row 3: IonQ.}  IonQ exposes native gates via API since 2022, with phase
tracked at the pulse level for floating-point
fidelity~\cite{ionqNative,ionqNativeAPI}; the Aria-class systems use the
M{\o}lmer--S{\o}rensen gate and the Forte-class systems use the ZZ gate, but
neither exposes arbitrary pulse waveforms.  Access is via direct API plus AWS
Braket and Google Cloud Marketplace integrations, essentially unchanged since
2022.  This row, with Quantinuum, establishes that trapped-ion vendors have
settled at native-gate-only access as a stable equilibrium that is neither
closing nor opening, and the absence of visible community pressure to change
suggests the equilibrium fits the use cases trapped-ion researchers prioritize.

\textbf{Row 4: Quantinuum.}  Quantinuum exposes a native gate set (parameterized
RZZ and fully entangling ZZ by default, with Rxxyyzz($\alpha,\beta,\gamma$) on
opt-in) accessible via standard compilation flags; pulse-level shaping is not
exposed publicly~\cite{quantinuumSystems}.  Access is via the Quantinuum Systems
API plus Microsoft Azure Quantum integration, with SDK license partial via TKET
(Apache 2.0) plus the proprietary direct client.  The structural role is shared
with row 3.

\textbf{Row 6: Rigetti.}  Quil-T provides mature pulse-level control with custom
waveforms, frames, sample-rate awareness, parameterized templates such as
\texttt{drag\_gaussian}, and direct waveform envelope
manipulation~\cite{rigettiQuilt,rigettiPyQuilDocs}.  pyQuil is Apache 2.0; QCS
plus AWS Braket plus Azure routing exposes the access through multiple cloud
platforms; Quil-T defines a stable abstraction over Rigetti's proprietary
control electronics.  This row is the second open-pulse counterexample in the
superconducting modality and shows that the ``mid-tier closes faster'' reading
does not survive contact with the catalog: Rigetti is one of the older
superconducting vendors and remains open at the pulse level.

\textbf{Row 8: QuEra.}  Bloqade (Julia and Python flavors) supports analog mode
on QuEra's cloud-accessible Aquila system, with the SDK fully open-source and
pointed directly at the hardware~\cite{queraBloqade,queraBloqadeBlog}.  SQUIN is
the digital DSL with pulse-level composability planned; Aquila access is via AWS
Braket; SDK license is Apache 2.0.  The open-by-design role at the SDK and pulse
layer for neutral atoms is shared with row 7.

\textbf{Row 9: Atom Computing.}  Atom Computing routes public access entirely
through Microsoft Azure Quantum integration; no public pulse-level SDK exists,
and the logical-qubit work on Phoenix and the upcoming Magne system (50 logical
qubits, 1{,}200 physical, 2027 delivery) goes through the Azure abstraction
layer~\cite{atomComputing2025,microsoftAtom2024}.  Four of the six axes grade
closed; the reproducibility cell grades limited on the best-effort posture of
the Azure integration, and the SDK license cell grades limited because the
Azure Quantum SDK exists but exposes Atom Computing only through gate-level
abstractions.  This is the
closed-via-cloud-abstraction case in the neutral-atom modality and shows that
institutional access patterns matter more than the underlying physics: Atom
Computing's hardware is in the same broad class as Pasqal and QuEra, but the
access posture is structurally opposite.

\textbf{Row 10: OQC.}  Oxford Quantum Circuits exposes OpenPulse on Lucy via AWS
Braket Pulse since 2022, with microwave-pulse control of manipulation and
readout at the OpenQASM 3 grammar level~\cite{oqcOpenPulse,awsBraketPulse}.
Access is via AWS Braket plus OQC private QCaaS; SDK license is open via the AWS
Braket SDK (Apache 2.0); the cloud-platform integration is the openness vehicle
here, not the closure mechanism.  This row is the cloud-platform counterexample
in the catalog: it shows that a cloud abstraction layer can be the openness
vehicle rather than the closure mechanism, depending on what the underlying
vendor chooses to expose through it.

\textbf{Row 11: Xanadu.}  Strawberry Fields and PennyLane are both Apache 2.0;
Borealis was the first photonic quantum computer publicly available via cloud
(AWS Braket, 2022), and Aurora was announced in 2025 as a 12-qubit modular
networked photonic computer~\cite{xanaduStrawberry,awsXanadu}.  Analog control
of squeezing parameters and beamsplitter networks is exposed through Strawberry
Fields; access is via AWS Braket plus Xanadu Cloud.  This row is the
open-by-design photonic case and pairs with PsiQuantum (closed by design, row
12) and ORCA (mid-tier, row 13) to make the photonic modality the cleanest
internal control for the modality-does-not-predict-openness finding.

\textbf{Row 13: ORCA Computing.}  ORCA's PT-1 (exploration) and PT-2 (utility)
systems use a PyTorch-based SDK with NVIDIA CUDA-Q integration; cloud access is
via institutional partnerships (NQCC, DOE labs) rather than a public cloud
queue, and the level of pulse-level or analog-parameter control exposed through
the SDK is not fully documented in publicly accessible sources as of the
cutoff~\cite{orcaTech}.  The ``see notes'' cells on the affected axes reflect
the limited public surface, and the per-row commentary substitutes for catalog
cells where the public documentation does not support a clean rubric grade.
This row is the mid-tier photonic case any single-cause modality argument has to
engage with: ORCA is photonic but neither open-by-design like Xanadu nor
closed-by-design like PsiQuantum, and the difference traces to commercial
strategy and target market rather than to the underlying physics.

\section{What the Application Shows}
\label{sec:results}

Five patterns fall out of Table~\ref{tab:catalog}.  Each is documented by
specific rows; the patterns are read off the catalog rather than imposed on it.

\textbf{The trajectory at the largest cloud platforms is closing.}  IBM Quantum
(row 1) deprecated Qiskit Pulse in November 2024 and removed pulse-level control
from all production QPUs on February 3,
2025~\cite{ibmPulseMigration,qiskitDeprecate13063}.  Google Quantum AI (row 2)
has never exposed raw pulse access on Sycamore through the public Cirq API;
calibration metrics are fetchable via \texttt{cirq\_google.Engine} only by
collaboration users, and waveform control is not exposed at
all~\cite{googleCirqCalib,googleCirqDevices}.  Atom Computing (row 9),
routed through the Azure Quantum abstraction, exposes no public pulse-level
SDK~\cite{atomComputing2025,microsoftAtom2024}.  PsiQuantum (row 12) has no
public access at any layer~\cite{psiquantumOmega,psiquantumConstruct}.  The four
largest cloud-platform integrations or stealth fault-tolerant programs all grade
closed on Axis 1.

\textbf{The mid-tier superconducting story is the opposite.}  IQM (row 5) ships
Pulla as a client-side Python library on PyPI with full step-by-step compilation
transparency from circuit to pulse schedule~\cite{iqmPulla,iqmPullaPyPI}.
Rigetti (row 6) maintains Quil-T as a mature pulse-level extension to Quil, with
custom waveforms, frames, sample-rate awareness, and direct waveform envelope
manipulation~\cite{rigettiQuilt,rigettiPyQuilDocs}.  OQC (row 10) exposes
OpenPulse on Lucy via AWS Braket Pulse at the OpenQASM 3 grammar
level~\cite{oqcOpenPulse,awsBraketPulse}.  Three commercially funded
superconducting vendors at non-trivial scale are open at the pulse layer; the
catalog's strongest rebuke to a pure ``the field is closing'' narrative is that
the mid-tier moved in the opposite direction over the same three-year period.

\textbf{Trapped-ion vendors are stable at native-gate-only access.}  IonQ
(row 3) and Quantinuum (row 4) both expose native gate access via API, with
phase tracked at the pulse level for floating-point
fidelity~\cite{ionqNative,ionqNativeAPI,quantinuumSystems}.  Neither exposes
arbitrary pulse waveforms.  The posture has been stable for several years and is
not visibly trending in either direction; it is a different equilibrium, and it
suggests the modality has settled on a level of abstraction the vendors consider
commercially appropriate and that researchers have not pushed back hard enough
to change.

\textbf{Neutral atom is bimodal.}  Pasqal (row 7) ships Pulser as an Apache-2.0
Python package designed at the pulse layer and open by design from the
start~\cite{pasqalPulser}.  QuEra (row 8) ships Bloqade in Julia and Python
flavors for analog mode on Aquila, with SQUIN as the digital DSL and pulse-level
composability planned~\cite{queraBloqade,queraBloqadeBlog}.  Atom Computing
(row 9) routes public access entirely through Azure Quantum and exposes no
public pulse-level SDK~\cite{atomComputing2025}.  The split is institutional
rather than technical: the same physical modality supports maximally open SDKs
in two cases and a closed-via-cloud-abstraction posture in the third.

\textbf{Photonic is split three ways.}  Xanadu (row 11) is open by design at the
SDK and cloud-access layer, with Strawberry Fields and PennyLane both Apache 2.0
and Borealis available on AWS Braket since
2022~\cite{xanaduStrawberry,awsXanadu}.  PsiQuantum (row 12) is closed by design
at all layers, with the Omega chipset announced February 2025 and Construct as a
non-public algorithm-design platform~\cite{psiquantumOmega,psiquantumConstruct}.
ORCA Computing (row 13) sits in the middle, accessed through institutional
partnerships with a PyTorch-based SDK whose pulse-level surface is not fully
documented in public sources~\cite{orcaTech}.  The photonic modality alone
produces three openness postures, which makes the row trio a useful internal
control on any single-cause explanation.

\subsection{The Refined Finding: Modality Alone Does Not Predict Openness}

The cleanest reading of Table~\ref{tab:catalog} is that \textbf{modality alone
does not predict openness.}  Commercial position, scale of cloud user base, and
time of stack design predict more.  Vendors that started open tend to stay open:
Pulser, Bloqade, and Strawberry Fields were designed at the pulse or analog
layer from the beginning and have remained that way.  Vendors that grow large
cloud user bases tend to close: IBM is the cleanest example.  Stealth
fault-tolerant programs start closed: PsiQuantum has been closed at every layer
since founding.  The same physical substrate can support either posture, and the
catalog's rows make that visible per modality.  Superconducting includes IBM
closed and IQM open; neutral atom includes Atom closed and Pasqal open; photonic
includes PsiQuantum closed and Xanadu open.  The commercial and historical
determinants of openness are doing the work that modality cannot.

\section{Demonstration: A Concrete Harm}
\label{sec:demo}

The catalog establishes that pulse-level access is closed on IBM and open on
Rigetti.  This section demonstrates that the difference is not bookkeeping: it
determines whether a specific published experiment can be reproduced.  The
demonstration is a reproduction-access audit, not a result-reproduction.  The
question is whether the access a methods paper depended on still exists, and
where the rubric predicted each break, not whether the re-run reproduces the
original numbers.  The delivered demonstration is a reproduction-access audit of
five pre-2025 pulse-level methods papers, the failure side checked against the
current IBM SDK, and a client-side structural port of the audit's selected
target to Rigetti Quil-T; the port's outcome is a second closure mechanism,
gating by access channel, distinct from IBM's removal.  Hardware execution of
the ported program is the one step left out of scope.

\subsection{Reproduction-Access Audit of Five Pulse-Level Papers}

Five pre-2025 methods papers, each of which depended on public pulse-level access
on IBM, are audited along three questions: which rubric axes the methodology
required, whether those axes are open on the original platform at the May 1, 2026
cutoff, and which currently open platform could host an equivalent re-run.  The
five span the range of pulse-level dependence: the framework paper that
introduced Qiskit Pulse~\cite{alexanderQiskitPulse2020}; sample-level pulse
compilation bootstrapped from gate calibrations~\cite{gokhaleOpenPulse2020};
in-situ closed-loop optimal control of each waveform
sample~\cite{werninghausLeakage2021}; scaled cross-resonance pulses for
two-qubit transpilation~\cite{earnestPulseEfficient2021}; and pulse-level
calibration of a non-Clifford controlled-$S$ gate~\cite{garionNonClifford2021}.

\begin{table*}[t]
\caption{Reproduction-access audit of five pre-2025 IBM pulse-level methods
papers.  Every paper required IBM Axis 1 at level open; none is re-runnable at
the pulse layer on IBM at the cutoff.  The rightmost column names the
currently-open platform that could host an equivalent, with the qualifier the
audit surfaced.}
\label{tab:audit}
\centering
\small
\begin{tabularx}{\textwidth}{@{}>{\hsize=.75\hsize}Y>{\hsize=1.0\hsize}Y>{\hsize=.75\hsize}Yc>{\hsize=1.5\hsize}Y@{}}
\toprule
Paper & Pulse dependency & Axes required & IBM now & Equivalent re-run platform \\
\midrule
Alexander et al. 2020~\cite{alexanderQiskitPulse2020} &
raw waveform synthesis, schedules & A1 open &
closed & Rigetti Quil-T, IQM Pulla, OQC OpenPulse (direct) \\
Gokhale et al. 2020~\cite{gokhaleOpenPulse2020} &
sample-level pulses from gate calibrations & A1 open, A2 partial &
closed & Rigetti, IQM, OQC (A1 open, A2 partial on each) \\
Werninghaus et al. 2021~\cite{werninghausLeakage2021} &
closed-loop optimization of each sample & A1 open, A4 open, A2 &
closed & none: A1 open on Rigetti/IQM/OQC but A4 is partial everywhere \\
Earnest et al. 2021~\cite{earnestPulseEfficient2021} &
scaled cross-resonance pulses & A1 open, A2 &
closed & OQC (fixed-frequency transmon); Rigetti/IQM differ in gate architecture \\
Garion et al. 2021~\cite{garionNonClifford2021} &
pulse-level two-qubit gate calibration & A1 open, A2 &
closed & Rigetti, IQM, OQC \\
\bottomrule
\end{tabularx}
\end{table*}

All five papers required IBM Axis 1 at level open, and all five are currently
un-runnable at the pulse layer on IBM, because Axis 1 closed on February 3,
2025~\cite{ibmPulseMigration}.  The rubric predicts every one of these breaks
from a single cell: the same cell whose movement Section~\ref{sec:sensitivity}
dates.  The migration target, fractional gates plus Qiskit Dynamics, does not
restore sample-level waveform control, so the two sample-level papers
(Gokhale, Werninghaus) lose the layer their methodology operated on rather than
gaining a replacement for it.

The audit also surfaces two boundaries that a one-cell reading would miss, and
both sharpen what the rubric does and does not claim.  Axis 1 open is necessary
but not sufficient for a port: the cross-resonance method of
Earnest et al. depends on fixed-frequency transmons coupled by cross-resonance,
and an Axis-1-open platform of a different architecture (Rigetti and IQM both
use tunable couplers) exposes the waveform but cannot host the same gate
physics; OQC, a fixed-frequency transmon platform, is the closest match.  The
rubric measures access to the control layer, not equivalence of the underlying
hardware, and the audit makes that boundary explicit.  The hardest dependency
to satisfy anywhere is Axis 4: the in-situ closed-loop optimization of
Werninghaus et al. needs real-time scheduling that no surveyed vendor grades
open, so even with pulse access reopened, that class of experiment remains
out of reach on current commercial cloud platforms.  This is the strongest
single result of the audit: closure of one axis (IBM Axis 1) removes five
published methods from their origin platform, and the structure of the
remaining open platforms determines, axis by axis, which of the five can be
rehosted and which cannot.

\subsection{The Failure Side on IBM}
\label{sec:demo-ibm}

The audit's central claim, that the five papers are un-runnable at the pulse
layer on IBM, is checkable directly against the current public SDK without any
QPU time, because the construction fails before a job is ever submitted.  A
reproducible probe (\texttt{demo/ibm\_pulse\_probe.py} in the artifact) imports
the pulse constructs each paper depended on against Qiskit 2.4.1 and
qiskit-ibm-runtime 0.47.0, the public releases installed in June 2026, and
records the outcome.

The result is unambiguous.  Importing \texttt{qiskit.pulse} raises
\texttt{ModuleNotFoundError: No module named 'qiskit.pulse'}: the module the
framework paper introduced~\cite{alexanderQiskitPulse2020}, and on which the
other four built, no longer exists.  With it go \texttt{qiskit.pulse.library},
so the \texttt{GaussianSquare} envelope Earnest et al. scaled for
cross-resonance~\cite{earnestPulseEfficient2021} and the arbitrary-sample
\texttt{Waveform} that Gokhale et al. and Werninghaus et al. optimized
sample by sample~\cite{gokhaleOpenPulse2020,werninghausLeakage2021} cannot be
constructed at all.  \texttt{QuantumCircuit.add\_calibration}, the path by which
Garion et al. attached a pulse-calibrated controlled-$S$ schedule to a gate~%
\cite{garionNonClifford2021}, is also removed.  \texttt{QuantumCircuit} itself
remains: gate-level circuits still run, which is the point.  The layer the five
methods operated on is the layer that is gone, and the layer above it is
untouched.

The documented migration target is fractional gates, RZZ rotations at arbitrary
angle built into the instruction set, plus Qiskit Dynamics for
simulation~\cite{ibmPulseMigration}.  Neither restores sample-level waveform
synthesis or arbitrary pulse calibration to public-API users.  A fractional gate
is a parameterized native gate, not an editable envelope; it abstracts past the
waveform layer rather than reopening it.  The methods that lived at that layer
therefore lose it rather than gaining a replacement, which is what Axis 1 closed
records.

\subsection{The Port Side on Rigetti}
\label{sec:demo-rigetti}

Porting one paper to an Axis-1-open platform carries the demonstration from a
documentary claim toward a run result.  The audit revises the target choice:
Earnest et al. is the worst port candidate, not the best, because its
cross-resonance method does not survive the architecture change to Rigetti's
tunable couplers; an architecture-agnostic single-qubit pulse-shaping or
two-qubit calibration result, in the manner of Garion et al., is the right
target precisely because the audit shows it ports where the cross-resonance
method does not.  Rigetti's pulse layer is exposed through Quil-T, and the
catalog grades it open.  The port attempt below establishes that this openness
is real but channel-dependent: the same pulse access is reachable through one
access path and gated through another.  The current Rigetti device is
Cepheus-1-108Q.

\subsubsection{The Azure cloud-broker channel}

Cepheus-1-108Q is reachable through Azure Quantum: the target is available and
accepts native Quil (\texttt{rigetti.quil.v1}), confirmed against a paid
workspace in June 2026.  Constructing a Quil-T pulse program, however, requires
the device's pulse calibrations, the \texttt{DEFFRAME}, \texttt{DEFWAVEFORM},
and \texttt{DEFCAL} set that fixes each frame's hardware parameters.  Those
calibrations are served by Rigetti's Quantum Cloud Services (QCS), and fetching
them through the Azure-authenticated session fails: the QCS calibration call
returns a credential error because the Azure broker forwards job submission but
not QCS calibration access, and no QCS credential is present.  The precise,
rigorous claim is narrow: \emph{client-side construction of a valid Quil-T
program is blocked through the Azure channel without separate QCS credentials.}
We do not claim the QPU would reject a submission, since server-side translation
may inject calibrations; we claim only that the pulse layer the methodology
needs is not reachable for program construction through this channel.  This is a
second closure mechanism distinct from IBM's: IBM removed the pulse API
outright, whereas Rigetti's pulse layer exists but is gated by the access
channel.  An open axis can be closed in practice by the path a user takes to it,
which the rubric measures at the documentation surface and which this attempt
shows can diverge by channel.

\subsubsection{The QCS-direct channel and the structural port}

The controlled comparison is the same access attempted through Rigetti QCS
directly, the channel the catalog graded.  No QCS account was held during this
study, so the QCS-direct calibration fetch is reported at the boundary it
presents to a credential-less client rather than from inside an account: the
calibration set is served through a documented public API endpoint
(\texttt{get-quilt-calibrations})~\cite{rigettiQCSAPI}, the open-source
\texttt{qcs-sdk} client exposes the call, and invoking it without a QCS
credential fails at exactly that step with a credential error (no refresh
token configured).  The boundary is a sign-up requirement, not a removed
interface; it is the same boundary the Azure attempt hit from the other side,
where job submission is forwarded and the calibration call is not.

The structural port, the load-bearing half of the demonstration, does not
require crossing that boundary, and it is executed rather than planned.  The
target is the architecture-agnostic one the audit selects: a pulse-calibrated
controlled-$S$ gate, $\mathrm{CS}=\mathrm{CPHASE}(\pi/2)$, in the manner of
Garion et al.; on Rigetti the CPHASE family is native to the tunable-coupler
architecture, which is what lets this target port where cross-resonance does
not.  The port program (\texttt{demo/rigetti\_quilt\_port.py} in the artifact)
constructs the full Quil-T text client-side: five explicitly declared frames
(\texttt{DEFFRAME} with direction, frequency, hardware object, and sample
rate), a raw-sample single-qubit envelope (\texttt{DEFWAVEFORM}), the gate
calibration \texttt{DEFCAL CPHASE(pi/2) 0 1} as a flat-top
\texttt{erf\_square} pulse on the two-qubit frame with local
\texttt{SHIFT-PHASE} corrections on each qubit, and a kerneled readout
(\texttt{DEFCAL MEASURE} with \texttt{NONBLOCKING CAPTURE}).  Against pyQuil
4.17.0 and its quil-rs grammar implementation, the public parser Rigetti's own
tooling builds on~\cite{rigettiPyQuilDocs}, the program parses and round-trips
stably: five frames, one custom waveform, two gate calibrations, one measure
calibration.  Every pulse-level construct the original methodology used at the
layer IBM removed is expressible and machine-checkable on the Rigetti side
from public artifacts alone.

\begin{table}[t]
\caption{Construct mapping from the Qiskit Pulse surface the five audited
papers used to the Quil-T surface the port lands on.  Direct: same construct
under a different name.  Restructured: analogue exists, the program reshapes
around it.  Gated: analogue exists but its invocation requires a QCS
credential.  None: no public analogue.}
\label{tab:portmap}
\centering
\scriptsize
\setlength{\tabcolsep}{3pt}
\begin{tabularx}{\columnwidth}{@{}>{\hsize=.95\hsize}Y>{\hsize=1.05\hsize}Yl@{}}
\toprule
Qiskit Pulse construct & Quil-T analogue & Status \\
\midrule
\texttt{Waveform} (sample array) & \texttt{DEFWAVEFORM} & direct \\
\texttt{Drag} template & \texttt{drag\_gaussian} & direct \\
\texttt{GaussianSquare} template & \texttt{erf\_square}, \texttt{flat} & restr. \\
\texttt{Play} on \texttt{DriveChannel} & \texttt{PULSE} on \texttt{"rf"} frame & direct \\
\texttt{ShiftPhase}, \texttt{Set}/\texttt{ShiftFrequency} & \texttt{SHIFT-PHASE}, \texttt{SET-}/\texttt{SHIFT-FREQUENCY} & direct \\
\texttt{Delay} (\texttt{dt} samples) & \texttt{DELAY} (seconds) & restr. \\
\texttt{ControlChannel} (CR drive) & named two-qubit frame & restr. \\
\texttt{align\_*} builder contexts & \texttt{FENCE}, \texttt{NONBLOCKING} & restr. \\
\texttt{Acquire} to \texttt{MemorySlot} & \texttt{CAPTURE} with kernel & restr. \\
\texttt{add\_calibration} & \texttt{DEFCAL} & direct \\
\texttt{backend.defaults()} cal.\ set & \texttt{get-quilt-calibrations} & gated \\
meas.-conditioned in-sequence loop & none public (Axis 4) & none \\
\bottomrule
\end{tabularx}
\end{table}

Table~\ref{tab:portmap} records the construct-by-construct outcome.  Five
constructs map directly, five restructure (the frame model is explicit and
named where IBM's channel model was indexed; timing is in seconds rather than
samples; readout requires an explicit kernel), one is credential-gated, and
one has no public analogue.  The two boundaries the mapping surfaces are the
same two the audit found.  The gated row is the device calibration set, the
Quil-T analogue of the \texttt{backend.defaults()} bootstrap that
Gokhale et al. depended on: the port program substitutes explicitly declared
placeholder frame parameters, which is sufficient for structural validity and
marks precisely the data the gated call would supply.  The no-analogue row is
the measurement-conditioned real-time loop of Werninghaus et al., which is an
Axis 4 dependency no surveyed vendor satisfies publicly.  Carrying the port
from structural validity to a hardware run requires a QCS account for the
calibration fetch and paid QPU time for execution; both are out of scope here,
and neither changes the structural result: the pulse layer that closed on IBM
is constructible and checkable on Rigetti from public artifacts, up to the
calibration values that the access channel determines.

\section{Limitations and Alternative Interpretations}
\label{sec:steelman}

Eight alternative interpretations of the catalog readings deserve direct
treatment.  The pulse-access difficulty argument
(Section~\ref{sec:steel1}) and the consolidation argument
(Section~\ref{sec:steel6}) receive extended responses; the remaining six are
addressed in turn.

\subsection{Pulse access is genuinely hard to expose safely}
\label{sec:steel1}

Calibration drift makes raw pulse schedules brittle on a per-shot basis.
Hardware-protection guards are needed to prevent users from damaging cryogenic
systems with badly shaped envelopes.  The API surface of pulse-level control
creates a real support burden because every user is doing something different.
Documenting the control electronics correctly is engineering work that gets
done by hand, slowly.  These constraints are real, and exposing pulse-level
access to general cloud users is a hard product decision with genuine downside
risk.  IBM's stated rationale for the February 2025 deprecation makes this
defense in its own words: the most common use of pulse-level control was
building custom schedules to modify ECR or RX pulses for single- and two-qubit
rotations, and fractional gates offer similar capability through a higher-level
interface better suited for quantum advantage
applications~\cite{ibmPulseMigration}.

The engineering difficulty of safely exposing user-facing pulse access does not
justify closing the closed-loop feedback layer or the control-electronics
interface.  Those are separable concerns.  A vendor can choose not to expose
arbitrary pulse waveforms to general users and still publish the schema by which
its internal feedback controller talks to its electronics, the algorithms that
run in the FPGA gateware between measurement and control, the protocol for
swapping in third-party control electronics, and the calibration data as a
published time series~\cite{qubic2,openHardware2024}.  The IBM rationale is fair
as a product decision, but it does not address the research community whose work
depends on the less common cases: pulse-level dynamical decoupling sequences,
non-standard gate decompositions, novel feedback protocols, and the methods
papers the gate-level community ends up depending on a few years later.
Pulse-access difficulty argues for tiered access models with safety-aware
abstractions; it does not argue for closed sourcing of the control plane below
the user-facing API, and the conflation of the two is what this paper pushes
against most directly.

\subsection{Calibration IP is real commercial value}

Vendors invest substantial physics and engineering effort into calibration
routines, drift compensation, gate-set decomposition, and pulse shaping.  That
work is differentiating, and open-sourcing it would erode the advantage and
reduce the incentive to fund the work in the first place.

Calibration IP and calibration data are different objects.  Routines are IP; the
data those routines produce on a given device on a given day is not the same
thing, and publishing the data does not publish the routines.  A floor for
openness can ask for the data without asking for the routines, which is the same
posture most semiconductor foundries take with published process design kits
while keeping their internal characterization stack
proprietary~\cite{edwardsSkywater2020}.

\subsection{The field is too early to standardize}

Premature standardization locks in bad design choices.  Quantum hardware is
heterogeneous enough that a control-plane interface defined now would be wrong
in five years, so it is better to let the design space mature.

There is a difference between freezing an interface and documenting one.  The
catalog shows that several vendors already have mature, documented pulse-level
interfaces (Pulser, Bloqade, Quil-T, Pulla); the paper does not propose freezing
those into a single specification.  The interfaces are legible, citable, and
stable enough that experiments running through them can be reproduced.  The
Munich QDMI effort and the QIR Alliance both demonstrate that interface work is
happening anyway~\cite{munichQDMI,qirAlliance}; the question is whether it
happens openly with broad input or behind cloud SLAs.

\subsection{Most researchers don't need pulse access}

Cloud usage statistics across IBM Quantum, Azure Quantum, AWS Braket, and
similar platforms suggest the overwhelming majority of jobs are gate-level, so
closing pulse access affects a small minority and the cost-benefit calculation
is rational.

The population of users matters less than the population of papers.  The
pulse-level minority disproportionately produces the methods papers,
error-correction experiments, calibration studies, and benchmarking results the
gate-level majority depends on.  A field where the foundational papers cannot be
reproduced because the pulse-level access they used is no longer available is not
operating on a sound base.  The fraction of cloud jobs is the wrong denominator.

\subsection{Hardware specificity makes openness impractical}

Every superconducting fab is different; every ion-trap geometry is different;
every neutral-atom array has its own cooling and trapping geometry.  Pulses that
work on one device do not work on another, so open access at the pulse level is
not portable and the meaning of openness here is unclear.

Openness at the control plane was never about pulse-level portability.  It is
about access to the actual interface for one's device, with documentation
sufficient to reproduce results, calibration data sufficient to interpret them,
and electronics interfaces sufficient to understand what physical loop closed
when.  Hardware specificity is a reason to publish more documentation per
device, not less.

\subsection{The trajectory is consolidating, not closing}
\label{sec:steel6}

What reads as ``closing'' is consolidation around abstraction levels appropriate
to different audiences.  Cloud users want gate-level access and a queue that
returns answers quickly; pulse-level access can remain available to research
partners under separate agreements.  IBM's deprecation of Qiskit Pulse on cloud
QPUs is consistent with this read: the access has not vanished from the
universe, only from the public cloud tier.  Vendors are differentiating their
offerings across audience segments, which is normal commercial behavior and
which produces a better experience for each segment than a single one-size
interface would.

The consolidation read is honest about the commercial logic but evades the
reproducibility problem.  ``Available to research partners under separate
agreements'' is not the same as ``publicly accessible.''  A result that depended
on access available only to a privileged few cannot be reproduced by anyone
outside that circle, which is a property of closed access regardless of what the
vendor calls it.  The paper is about what is publicly citable, reproducible, and
auditable, and the catalog grades on that axis: a vendor that offers pulse
access only under NDA grades closed at the public layer, with a note that
private access exists.  The consolidation framing is also incomplete on the
timeline.  Vendors that moved over the last three years moved overwhelmingly
toward closing public access, not toward consolidating multiple tiers; if the
trajectory were consolidation, the catalog would show a fan-out of access levels
at large vendors, and what it shows is contraction.

\subsection{Open access creates safety and security risks}

Pulse-level APIs let users emit arbitrary microwave envelopes, and a malformed
pulse can damage cryogenic systems or saturate amplifiers downstream; the
attack surface pulse-level circuits open has been studied
directly~\cite{xuSzefer2024}.
Calibration data publication can leak side-channel information about device
topology, decoherence characteristics, and fab-process variation.  Open
electronics interfaces broaden the attack surface for denial-of-service and
resource exhaustion.  Vendor closure at the control plane is partly a
security-perimeter decision, and security-perimeter decisions are appropriately
made by vendors rather than by external commentators.

The security concerns are real but they argue for guarded openness, not full
closure.  Tiered access with safety-aware abstractions, sandboxed test
environments before pulse submission, rate limiting, and parameter range
validation are standard mitigations other engineering domains use to expose
risky interfaces safely.  The semiconductor PDK analogue is informative:
SkyWater's open PDK exposes process-design rules that could in principle be used
to design adversarial chips, and the response was layered access with
documentation-quality gates, not closure~\cite{edwardsSkywater2020,skywaterPDK}.
Calibration data side-channels are a more interesting concern, but the catalog
grades calibration publication separately from pulse access for exactly this
reason: a vendor can publish coarse-grained metrics ($T_1$, $T_2$, gate error
rates) without publishing the fine-grained device-fingerprint data that creates
real side-channel risk.

\subsection{Open access undercuts the funding model that pays for the hardware}

The field's progress depends on commercial vendors profiting; investors fund the
hardware capital expenditure on the expectation of returns; openness erodes the
moat that justifies the capital.  If the control plane is open, the remaining
monetization surface is the hardware itself, and public funding cannot replace
private capital at the scale current quantum hardware development requires, so an
argument that erodes private capital is an argument against the field's progress.

The funding-model argument has a real core but conflates two layers.  The
hardware itself remains a moat: no one is open-sourcing dilution refrigerators,
ion-trap geometries, photonic chip fabrication processes, or laser systems.  The
control plane is a thin layer of software and electronics interfaces on top of
that moat, and the catalog shows openness at this layer is compatible with
running a commercial business.  IQM and Rigetti both run commercial businesses
with open pulse-level access (Pulla, Quil-T) and have not visibly suffered for
it; Pasqal and QuEra both run commercial businesses with open neutral-atom SDKs
(Pulser, Bloqade) and have not visibly suffered for it.  The argument would be
stronger if the closed layer were the moat layer; in practice they are not.  A
version focused on differentiation specifically (vendors need \emph{something}
proprietary to differentiate) is more honest, and the response there is that
hardware, calibration routines, fabrication processes, and product polish all
remain proprietary; the control plane is not the only differentiation surface,
and arguably not even the most important one.

\section{Lessons from Open-Infrastructure Precedents}
\label{sec:analogues}

Three precedents from open-infrastructure transitions in adjacent fields offer
lessons for the control-plane case.\footnote{We use Linux, LLVM, and the
SkyWater PDK rather than Python or GCC, which are too easily dismissed as
cherry-picked; the SkyWater PDK in particular is a closer structural fit to the
control-plane question than either of the other two.}  The three do not predict
the same outcome for quantum, but they show that the structural pattern, a
closed bottom-of-stack interface under pressure to open, is not new.

\textbf{Linux.}  The right framing is not ``open beat closed'' but ``open won at
the layer everyone insisted was too low-level to matter''~\cite{raymondBazaar}.
In 1991 the conventional wisdom said operating-system kernels were not where open
source could compete; the bet that paid off was at the layer with the most
direct hardware contact, the highest engineering complexity, and the least
obvious commercial appeal.  \emph{Lesson: the layer dismissed as too low-level to
open is the layer where open access compounds the most.}

\textbf{LLVM.}  An open compiler middle-end made hardware-aware research possible
across vendors~\cite{lattnerLLVM2004}.  LLVM started in 2000 as a research
project at the University of Illinois Urbana--Champaign and grew into a substrate
that Clang, Swift, Rust, Julia, Mojo, and effectively every production JIT now
share.  \emph{Lesson: an open layer where vendor specialization lives below the
line is what makes new hardware addressable by upstream tools.}

\textbf{SkyWater open PDK.}  The SkyWater 130nm PDK opened in 2020 a previously
NDA-locked layer of the semiconductor design stack, letting startups, academic
groups, and student projects design real silicon without proprietary access
agreements~\cite{skywaterPDK,edwardsSkywater2020}.  The control-plane analogy is
closer here than for Linux or LLVM in one specific way: the layer that opened
was the one where the hardware vendor's IP sat closest to the user, and the
opening was a deliberate commercial decision with measurable downstream effects
(Tiny Tapeout, OpenROAD adoption, the broader open-EDA flowering).  \emph{Lesson:
opening the layer where vendor IP sits closest to the user makes adjacent layers
more useful, which is the same dynamic that would follow from opening the control
plane.}

\section{Implications for Minimally Open Access}
\label{sec:floor}

This section identifies what minimally open access at the control plane would
require.  The framing is what the field needs, not what someone should build:
no architecture, runtime, stack shape, or reference implementation is proposed.
Four floor elements are named at the schema-shape level; each describes what
would have to be true for the element to count as met, without proposing a
specific format, library, or institution to maintain it.

\textbf{Open pulse APIs.}  Public-API users have a documented interface to pulse
synthesis on the hardware they are running on.  The interface is stable enough
that an experiment written against it last year still runs against it today, or
has a documented migration path to a successor.  Pulse-level access is not the
same as raw waveform exposure; what counts is that the user can specify
pulse-level properties of their experiment and have them faithfully executed.
Whether the interface is OpenPulse, Quil-T, Pulser, Pulla, or something else is
not the relevant question; that it exists, is documented, and is stable, is.

\textbf{Published calibration schemas.}  Per-device calibration data is
machine-readable, dated, retained for the operational lifetime of the device,
and covers at minimum the parameters needed to reproduce a published result.
Schema definitions are public so that consumers know what fields to expect and
what their semantics are.  Publication cadence is documented, even if the cadence
is per-job rather than periodic, and historical data is accessible to researchers
who need to interpret a result published months or years ago.

\textbf{Documented control electronics interfaces.}  The schema between SDK and
control electronics is published with enough detail that a third-party
electronics vendor could in principle build a compatible backend.  This does not
require open hardware; it requires that the interface between the vendor's
software and the vendor's hardware be a documented contract rather than an opaque
internal coupling.  Reference implementations such as QubiC~\cite{qubic2}
demonstrate that the layer can be built openly by national labs even when
commercial vendors are not yet ready to publish their own.

\textbf{Reproducibility guarantees.}  Returned results carry provenance metadata
sufficient to identify the calibration snapshot, the hardware revision, the SDK
version, and the queue tier under which the job ran.  Calibration is versioned
with stable identifiers.  The vendor publishes a commitment about how long a
given hardware revision is held stable, even if the commitment is short, so that
researchers can plan experiments whose validity depends on access to that
revision.

\subsection{What OQD is Already Building}

Open Quantum Design is one shape of the floor: a non-profit foundation building
reference open hardware in the trapped-ion modality, with partnerships across
academia and the open-source quantum community~\cite{oqdAnnouncement}.  This
paper does not call for another foundation of OQD's shape, but for broad vendor
adoption of the practices the floor names.  The two efforts are complementary:
OQD demonstrates that the floor is achievable in principle, and commercial
vendor adoption would demonstrate that it is achievable in the platforms most
researchers actually use.  Neither substitutes for the other.

\section{Conclusion}
\label{sec:conclusion}

This paper proposes a six-axis rubric for measuring control-plane openness, the
layer between gate-level circuit specification and physical control electronics,
validates it through a blinded re-grading stability test and a time-point
sensitivity analysis, and applies it to thirteen commercial vendors as the
rubric's first application.  Public access to this layer has bifurcated.  IBM's removal of
pulse-level control from all production QPUs on February 3, 2025 is the most
consequential single event in the period under review and the load-bearing case
for the closing-trajectory reading.  Mid-tier superconducting vendors (IQM,
Rigetti, OQC), the more open neutral-atom platforms (Pasqal, QuEra), and one
photonic vendor (Xanadu) have moved in the opposite direction.  Trapped-ion
vendors have held steady at native-gate-only access.  The photonic modality is
split three ways and demonstrates that modality alone does not predict openness.

The three harms named in Section~\ref{sec:harms} are concrete, not
hypothetical.  Reproducibility has been measurably eroded for IBM hardware as of
February 2025, and the demonstration in Section~\ref{sec:demo} shows the erosion
on a specific experiment.  Hardware-aware research at the pulse layer is gated by
which vendor a researcher chooses, and the choice is increasingly constrained on
the largest cloud platforms.  Cross-vendor benchmarking at the pulse layer is
structurally impossible across many of the catalog's rows because the underlying
interfaces differ in kind, not degree.

The catalog's strongest implication is not the closing trajectory itself but the
asymmetry of the closing.  Vendors that started open have stayed open; vendors
that grew large cloud user bases have closed; stealth fault-tolerant programs
started closed.  The mechanism is commercial position and historical inflection
point, not modality or technical necessity.  The floor sketched in
Section~\ref{sec:floor} is not difficult to meet on the technical side; the
elements are met already by vendors at smaller scale.  What changes is whether
the largest platforms decide that meeting the floor is part of running a quantum
cloud service.

We make no recommendation about how the floor should be enforced.  We do invite
three reactions.  Vendors who already meet the floor have a citable description
of what they are doing right and a place to point when asked to justify their
interface choices.  Funders considering whether to make openness a precondition
for grant solicitations have a vendor-by-vendor reference for what openness
currently looks like in the commercial landscape.  Researchers writing methods
papers have a citable artifact for the related-work paragraphs that describe
their hardware platform's access model.

The catalog readings will change; the rubric is the contribution that survives
them.  The catalog is released as a separate machine-readable artifact under
CC-BY-4.0 with source URLs and accessed-on dates per cell, meant as a living
dataset that other researchers can extend, correct, and disagree with on a
per-cell basis.  The rubric is the instrument that lets them do so consistently.

\section*{Disclosures}

During the grading period the author was developing an open-source pulse-control
software project that used one of the vendor stacks surveyed here (IQM) as its
primary integration target; that project has since been discontinued.  No
claim in this paper depends on or references it, and the IQM row is
graded only on public documentation.  The author was an undergraduate student in
the Department of Physical Sciences at Embry--Riddle Aeronautical University at
the time of writing, and is to participate in the Los Alamos National Laboratory
quantum computing summer school program for 2026.

\section*{Tool Use Disclosure}

The author used Claude and Codex to assist in literature search, catalog
drafting, and \LaTeX{} compilation.  The blinded re-grading pass of
Section~\ref{sec:regrading} was performed by an AI assistant in a fresh
session, operating only on the masked evidence pack described there (the
rubric level definitions, per-cell cited URLs, and recorded verbatim quotes,
with every recorded grade stripped) and on live fetches of the cited pages;
the pack, the re-grade output, and the comparison script ship with the
catalog artifact so the pass can be audited or repeated.  Every cell in the
catalog traces to a public source documented in the per-vendor catalog notes
files in the project repository.  No claim in the paper rests on a source the
author has not personally verified.

\section*{Catalog Availability}

The full machine-readable catalog, including source URLs and accessed-on dates
per cell, is available on Zenodo at
\url{https://doi.org/10.5281/zenodo.20163276}.  The catalog is licensed under
CC-BY-4.0.  The artifact also carries the re-grading protocol files of
Section~\ref{sec:regrading} (the masked evidence pack, the script that builds
it, the re-grade output, and the comparison script with its output), the
dated per-cell corrections, and the demonstration probes of
Section~\ref{sec:demo} (\texttt{demo/ibm\_pulse\_probe.py} and
\texttt{demo/rigetti\_quilt\_port.py}).

\bibliographystyle{IEEEtran}
\bibliography{refs}

\end{document}